\documentclass[preprint,showpacs,amssymb,aps,eqsecnum]{revtex4}
\usepackage{latexsym,hyperref}

\newcommand{\ket}[1]{\mbox{$|q_{#1},t_{#1}\rangle$}}
\newcommand{\bra}[1]{\mbox{$\langle q_{#1},t_{#1}|$}}

\newcommand{\amp}{\mbox{$\langle q_2,t_2|q_1,t_1\rangle$}}
\newcommand{\ampJ}{\mbox{$\langle q_2,t_2|q_1,t_1\rangle\lbrack J\rbrack$}}
\newcommand{\J}{\mbox{$\lbrack J\rbrack$}}

\begin{document}

\title{The Schwinger Action Principle and the
Feynman Path Integral for Quantum Mechanics in Curved Space  }

\author{David J. Toms}
\email{d.j.toms@newcastle.ac.uk}
\homepage{http://www.staff.ncl.ac.uk/d.j.toms/}
\affiliation{School of Mathematics and Statistics, University of Newcastle Upon Tyne,\\
Newcastle Upon Tyne, United Kingdom NE1 7RU}

\date{\today}

\begin{abstract}
The Feynman path integral approach to quantum mechanics is
examined in the case where the configuration space is curved. It
is shown how the ambiguity that is present in the choice of path
integral measure may be resolved if, in addition to general
covariance, the path integral is also required to be consistent
with the Schwinger action principle. On this basis it is argued
that in addition to the natural volume element associated with the
curved space, there should be a factor of the Van Vleck-Morette
determinant present. This agrees with the conclusion of an
approach based on the link between the path integral and
stochastic differential equations.
\end{abstract}

\pacs{04.62.+v,03.65.Ca,03.65.-w,03.70.+k}
\maketitle

\section{\label{sec1}Introduction}

There is a large body of work on the Feynman path
integral~\cite{Feynman48} approach to quantum mechanics on a
curved background, and the Schr\"{o}dinger equation which results
from the path integral. (See some of the references described
below.) It is widely known that, depending upon what is chosen for
the measure in the configuration space path integral, there is a
multiple of the scalar curvature to be added to the canonical
Hamiltonian in the resulting Schr\"{o}dinger equation. This is
demonstrated explicitly by Parker~\cite{Parker} who considers
including an arbitrary power of the Van
Vleck-Morette~\cite{VanVleck,Morette} determinant
$\Delta^p(x,x')$, where $p$ is an arbitrary real number, in the
measure in addition to the natural curved space volume element.
The choice $p = 0$, which corresponds to choosing the natural
volume element for the path integral measure, and $p = 1/2$, which
is motivated by the WKB approximation, were examined originally by
DeWitt~\cite{DeWitt57}. The choice $p = 1$ was shown by
Parker~\cite{Parker} to lead to no additional curvature
modification of the canonical Hamiltonian. For any value of $p$
the path integral measure is invariant under an arbitrary change
of coordinates for the curved space. General covariance alone of
the configuration space path integral does not prescribe a unique
measure.

It might be thought that this lack of uniqueness in the path
integral approach to quantum mechanics in curved space disappears
if a phase space path integral is used instead of one in
configuration space. In this case there is a unique choice of
measure; namely, the Liouville measure. However,
Kucha\v{r}~\cite{Kuchar} has shown in a very lucid paper that the
lack of uniqueness associated with the configuration space path
integral measure reappears in phase space as the lack of
uniqueness of the skeletonization of the action. (As discussed by
DeWitt~\cite{DeWitt57} Hamilton-Jacobi theory leads to a covariant
skeletonization in configuration space.) Lack of uniqueness in the
quantum theory is therefore inherent in the curved space Feynman
path integral, and cannot be eliminated solely on the grounds of
general covariance.

If this lack of uniqueness is to be eliminated from the Feynman
path integral in curved space, the only possibility is to impose
some further property in addition to general covariance under a
change of coordinates in the curved space. One example of an extra
property done within the minisuperspace approach to quantum
cosmology is to demand conformal invariance~\cite{Halliwell,Moss}.
However, this seems rather specific to quantum cosmology, and not
very compelling as a general principle, since there is no reason
to believe that conformal invariance is a fundamental symmetry of
nature. Another more general principle, based on the relationship
between the Feynman path integral and stochastic differential
equations, was given in Ref.~\cite{stochastic}. This corresponds
to the choice $p = 1$, and hence gives no curvature modification
to the canonical Schr\"{o}dinger equation. As we will see, the
approach adopted in the present paper supports the conclusion of
Ref.~\cite{stochastic}.

We will examine what happens if the Feynman path integral is
required to be equivalent to the Schwinger action
principle~\cite{SAP1,SAP2}. It was known almost from its
inception, that for flat spaces the Schwinger action principle is
completely equivalent to the path integral~\cite{Burton,Symanzik}.
This equivalence holds equally well for quantum field theory,
which is usually regarded as involving a flat configuration space.
(This is how DeWitt~\cite{DeWitt64}, for example, arrives at the
path integral.) In this paper, I wish to show how the Schwinger
action principle may be used to fix the measure in the Feynman
path integral for quantum mechanics on a curved space. If the
notation is interpreted in the spirit of DeWitt's~\cite{DeWitt64}
condensed notation, then the result holds equally well for quantum
field theory with a curved configuration space. (We will comment
briefly on this in Sec.~\ref{sec4}.) It will be shown that
equivalence between the Schwinger action principle and the Feynman
path integral is only achieved if there is a single factor of the
Van Vleck-Morette determinant in the measure, (i.e. $p = 1$
above.) This is completely consistent with the result of
Ref.~\cite{stochastic} which was based on totally different
reasoning.

\section{\label{sec2}The Schwinger action principle}

Let $q^i$ denote a set of local coordinates on some manifold $M$.
The classical motion of a particle moving on $M$ is given by
$q^i(t)$ where $q^i(t)$ is a solution to the Euler-Lagrange
equations. We will adopt a configuration space rather than a phase
space approach. In quantum mechanics we are interested in
computing the transition amplitude $\amp$ where $\ket{1}$
represents the quantum state at time $t_1$, and $\ket{2}$
represents the state at time $t_2\ge t_1$. These states are chosen
to be eigenstates of the position operator $\hat{q}^i$:
\begin{equation}
\hat{q}^i\ket{\alpha}={q}^i(t_\alpha)\ket{\alpha}\ \ (\alpha=1,2)
\;.\label{2.1}
\end{equation}
The Schwinger action principle~\cite{SAP1,SAP2} states that
\begin{equation}
\delta\amp=\frac{i}{\hbar}\bra{2}\delta S\ket{1}\;,\label{2.2}
\end{equation}
where $S$ represents the action obtained by the replacement of
$q^i$ in the action for the classical theory with $\hat{q}^i$,
along with an operator ordering which leads to $S$ being
self-adjoint. $\delta$ in Eq.~(\ref{2.2}) represents any possible
variation, including variations with respect to the times
$t_1,t_2$, the dynamical variables $q^i$, or the structure of the
Lagrangian. The variations of the dynamical variables $\delta q^i$
will be chosen to be c-numbers, appropriate to bosonic theories.
This choice was also made in flat space by Schwinger~\cite{SAP1},
and in curved space by Kawai~\cite{Kawai1,Kawai2}. The case of
fermionic variables will not be considered here.

Suppose that we add a source term to the action. Normally this is
done so that differentiation with respect to the source generates
the $n$-point functions of the theory. If the space is flat, this
may be accomplished by simply taking
\begin{equation}
S_J\lbrack q\rbrack = S\lbrack q\rbrack +
\int\limits_{t_1}^{t_2}dt\;J_i(t)q^i(t)\;,\label{2.3}
\end{equation}
where $S\lbrack q\rbrack$ is the original action, and $J_i(t)$ is
an external source which is turned on at time $t_1$ and off at
time $t_2$. However, on a curved space where $q^i$ are coordinates
rather than vectors, the addition of the source term in
Eq.~(\ref{2.3}) does not result in a covariant expression. This is
also the case in a flat space if the coordinates are chosen to be
curvilinear rather than Cartesian.

The covariant generalization of Eq.~(\ref{2.3}) is obtained as
follows. First, it may be noted that if the field space is flat,
the same classical theory is obtained from
\begin{equation}
S_J\lbrack q,q_\ast\rbrack = S\lbrack q\rbrack +
\int\limits_{t_1}^{t_2}dt\;J_i(t)\left(q^i(t)-
q_\ast^i(t)\right)\;,\label{2.4}
\end{equation}
as from as from Eq.~(\ref{2.3}), where the coordinates are assumed
to be Cartesian, and where $q_\ast^i$ is regarded as a fixed point
in the configuration space $M$. ($q_\ast^i$ plays the role that
the background field~\cite{DeWitt64} does in quantum field
theory.) The coordinate difference $(q^i - q_\ast^i)$ then
represents a vector which connects the fixed reference point
$q_\ast^i$ to the point $q^i$. Equivalently, $(q^i - q_\ast^i)$
represents the tangent vector to the geodesic connecting
$q_\ast^i$ to $q^i$, which for $M$ flat is just a straight line
segment. This indicates that the natural replacement for the
coordinate difference $(q^i - q_\ast^i)$ in a general space $M$ is
just the tangent vector at $q_\ast$ to the geodesic connecting
$q_\ast^i$ to $q^i$. One way of introducing this tangent vector is
by means of the geodetic interval $\sigma(q_\ast;q)$. (See
Refs.~\cite{DeWitt64,Ruse,Synge}.) By definition,
\begin{equation}
\sigma^i(q_\ast;q)=\frac{1}{2}\ell^2(q_\ast;q)\;,\label{2.5}
\end{equation}
where $\ell(q_\ast;q)$ is the length of the geodesic connecting
$q_\ast$ to $q$. The tangent vector to the geodesic at $q_\ast$ is
\begin{equation}
\sigma^i(q_\ast;q)=g^{ij}(q_\ast)\frac{\partial}{\partial
q_\ast^j} \sigma(q_\ast;q)\;,\label{2.6}
\end{equation}
where $M$ is assumed to have a metric tensor $g_{ij}$. If $M$ is
flat, and $q^i$ are Cartesian coordinates (so that
$g_{ij}=\delta_{ij}$), then
\begin{equation}
\sigma^i(q_\ast;q)=-(q^i-q_\ast^i)\;.\label{2.7}
\end{equation}
The natural replacement for $(q^i-q_\ast^i)$ in Eq.~(\ref{2.4}) is
therefore $-\sigma^i(q_\ast; q)$ resulting in
\begin{equation}
S_J\lbrack q, q_\ast\rbrack = S\lbrack q\rbrack -
\int\limits_{t_1}^{t_2}dt\;J_i(t)\sigma^i(q_\ast;q)\;.\label{2.8}
\end{equation}
$\sigma^i(q_\ast;q)$ transforms like a vector under a change of
coordinates $q_\ast$, and as a scalar under a change of
coordinates $q$. If the source $J_i(t)$ is required to transform
like a covariant vector at $q_\ast$, and be independent of $q$,
then Eq.~(\ref{2.8}) is a completely covariant definition. It is
important that $J_i(t)$ be independent of $q$ if it is to fulfill
its role as an external source. If $M$ is flat, but $q^i$ are not
Cartesian coordinates, it is possible to derive Eq.~(\ref{2.8})
from Eq.~(\ref{2.4}) using the approach of Ref.~\cite{Ellicott}.

It proves convenient to adopt condensed notation~\cite{DeWitt64}
at this stage, and to write Eq.~(\ref{2.8}) as
\begin{equation}
S_J\lbrack q,q_\ast\rbrack=S\lbrack
q\rbrack-J_i\sigma^i(q_\ast;q)\;.\label{2.9}
\end{equation}
where the index $i$ is now understood to include the time label,
and a repeated index includes integration over time. Instead of
regarding $S_J\lbrack q,q_\ast\rbrack$ as a functional of
$q,q_\ast$, it is convenient to regard it instead as a functional
$\tilde{S}_J\lbrack q_\ast;\sigma^i(q_\ast;q)\rbrack$ which is
defined using the covariant Taylor expansion~\cite{Ruse}
\begin{equation}
S\lbrack q\rbrack=\sum_{n=0}^{\infty}\frac{(-1)^n}{n!}
S_{;(i_1,\cdots i_n)} \lbrack q_\ast\rbrack \sigma^i_1(q_\ast;q)
\cdots \sigma^i_n(q_\ast;q) \label{2.10}
\end{equation}
in Eq.~(\ref{2.9}). (The semicolon denotes the usual covariant
derivative using the Christoffel connection constructed from the
metric $G_{ij}$ on $M$.)

Let $\ampJ$ be the transition amplitude for the theory with action
$S_J\lbrack q;q_\ast\rbrack$ in Eq.~(\ref{2.9}). The Schwinger
action principle gives
\begin{equation}
\delta\ampJ=\frac{i}{\hbar}\bra{2}\delta S_J\ket{1}\J \;.\label{2.11}
\end{equation}
If the variation is taken to be one that is with respect to the
dynamical variables $q^i$ leaving the values fixed at times $t_1$
and $t_2$, then the amplitude will not change under the variation,
and we have
\begin{equation}
0=\bra{2}\delta S_J\ket{1}\J \;,\label{2.12}
\end{equation}
from which the equation of motion may be inferred:
\begin{equation}
\frac{\delta\tilde{S}_J}{\delta\sigma^i}-J_i=0\;.\label{2.13}
\end{equation}

Since we may regard $\ampJ$ a functional of the source $J_i$, it
may be expanded in a Taylor series about $J_i = 0$:
\begin{equation}
\ampJ=\left.\sum_{n=0}^{\infty}\frac{1}{n!}J_{i_1}\cdots J_{i_n}\,
\frac{\delta^n\ampJ}{\delta J_{i_1}\cdots \delta
J_{i_n}}\right|_{J=0} \;.\label{2.14}
\end{equation}
We will now evaluate the $n^{\rm th}$ derivative of the amplitude
which occurs in Eq.~(\ref{2.14}) using the Schwinger action
principle. The method is just that used originally by
Schwinger~\cite{SAP1}.

Suppose that the variation in Eq.~(\ref{2.11}) is one with respect
to the external source $J_i$. Since the dependence of the action
on the source is given in Eq.~(\ref{2.9}), we have
\begin{equation}
\frac{\delta\ampJ}{\delta J_i}=-\frac{i}{\hbar}\bra{2}
\sigma^i(q_\ast;q) \ket{1}\J \;.\label{2.15}
\end{equation}
If we now perform a further variation of Eq.~(\ref{2.15}) with
respect to the source, we have
\begin{equation}
\delta\frac{\delta\ampJ}{\delta J_i}=-\frac{i}{\hbar}
\delta\bra{2} \sigma^i(q_\ast;q) \ket{1}\J \;.\label{2.16}
\end{equation}
In order to evaluate the right hand side of this expression,
insert unity in the form $1 =\int dv'|q',t'\rangle\langle q',t'|$
where $t_1 < t' < t_2$, and $dv' = d^n q'g^{1/2}(q')$ is the
invariant volume element on $M$. If the time included in the index
$i$ of $\sigma^i(q_\ast;q)$ lies to the past of $t'$, then we will
change the source only to the future of $t'$ and the past of
$t_2$. Assuming causal boundary conditions, $\delta\bra{2}
\sigma^i(q_\ast;q) \ket{1}\J$ cannot be affected by such a change
in the source. Thus,
\begin{eqnarray}
\delta\bra{2} \sigma^i \ket{1}\J &=&\int dv'\delta\langle
q_2,t_2|q',t'\rangle\J \langle q',t'|\sigma^i\ket{1}\J \nonumber\\
&=&-\frac{i}{\hbar}\int dv'\delta J_j\bra{2}\sigma^j|q',t'\rangle\J
\langle q',t'|\sigma^i\ket{1}\J \nonumber\\
&=&-\frac{i}{\hbar}\delta J_j\bra{2}\sigma^j\sigma^i\ket{1}\J \;,\label{2.17}
\end{eqnarray}
where we have dropped the argument $(q_\ast; q)$ on $\sigma^i$ and
$\sigma^j$ for brevity. Note that the time corresponding to the
condensed index $j$ lies to the future of that corresponding to
$i$.

Conversely, if $i$ lies to the future of $t'$, but to the past of
$t_2$, a similar argument shows that
\begin{equation}
\delta\bra{2} \sigma^i \ket{1}\J= -\frac{i}{\hbar}\delta
J_j\bra{2}\sigma^i\sigma^j\ket{1}\J \;.\label{2.18}
\end{equation}
Both situations in Eqs.~(\ref{2.17},\ref{2.18}) may be summarized
compactly by
\begin{equation}
\delta\bra{2} \sigma^i \ket{1}\J= -\frac{i}{\hbar}
\delta J_j\bra{2}T(\sigma^i\sigma^j)\ket{1}\J \;,\label{2.19}
\end{equation}
where $T$ is the chronological, or time ordering, symbol. It then
follows that
\begin{equation}
\frac{\delta^2\ampJ}{\delta J_{j}\delta J_{i}}=
\left(-\frac{i}{\hbar}\right)^2
\bra{2}T(\sigma^i\sigma^j)\ket{1}\J \;.\label{2.20}
\end{equation}
It is easily established by induction that
\begin{equation}
\frac{\delta^n\ampJ}{\delta J_{i_1}\cdots\delta J_{i_n}}=
\left(-\frac{i}{\hbar}\right)^n
\bra{2}T(\sigma^{i_1}\cdots\sigma^{i_n})\ket{1}\J \;.\label{2.21}
\end{equation}
Use of Eq.~(\ref{2.21}) in Eq.~(\ref{2.14}) leads to
\begin{eqnarray}
\ampJ&=&\sum_{n=0}^{\infty}\frac{1}{n!}
\left(-\frac{i}{\hbar}\right)^nJ_{i_1}\cdots J_{i_n}
\bra{2}T(\sigma^{i_1}\cdots\sigma^{i_n})\ket{1}
\lbrack J=0\rbrack\nonumber\\
&=&\bra{2}T\left\lbrace\exp\left(-\frac{i}{\hbar}J_i\sigma^i\right)
\right\rbrace \ket{1} \lbrack J=0\rbrack \;.\label{2.22}
\end{eqnarray}
(The exponential in the last line is understood to be defined in
terms of its Taylor series as in the preceding line.)

Define
\begin{equation}
E_i\left\lbrack q_\ast;\sigma^i(q_\ast;q)\right\rbrack=
\frac{\delta\tilde{S}}{\delta\sigma^i}\;,\label{2.23}
\end{equation}
so that the operator equation of motion Eq.~(\ref{2.13}) becomes
\begin{equation}
E_i\left\lbrack q_\ast;\sigma^i(q_\ast;q)\right
\rbrack=J_i\;.\label{2.24}
\end{equation}
We can view $E_i\left\lbrack q_\ast;\sigma^i(q_\ast;q)
\right\rbrack$ as defined in terms of the Taylor series obtained
by differentiating Eq.~(\ref{2.10}). Now consider
$\displaystyle{E_i\left\lbrack
q_\ast;-\frac{\hbar}{i}\frac{\delta}{\delta J_i}\right\rbrack}$
where $\sigma^i$ in the Taylor series for $E_i\left\lbrack
q_\ast;\sigma^i(q_\ast;q)\right\rbrack$ is replaced by
$-\frac{\hbar}{i}\frac{\delta}{\delta J_i}$. Using
Eq.~(\ref{2.22}), it is clear that
\begin{eqnarray}
E_i\left\lbrack q_\ast;-\frac{\hbar}{i}
\frac{\delta}{\delta J_i}\right\rbrack\ampJ&=&
\bra{2}T\left\lbrace E_i\left\lbrack q_\ast;
\sigma^i\right\rbrack \exp\left(-\frac{i}{\hbar}J_i\sigma^i\right
\rbrace \right) \ket{1} \lbrack J=0\rbrack\nonumber\\
&=&J_i\,\ampJ\;,\label{2.25}
\end{eqnarray}
noting Eqs.~(\ref{2.24},\ref{2.22}). This last result provides a
functional-differential equation for the amplitude $\ampJ$ which
has followed from the Schwinger action principle. Integration of
Eq.~(\ref{2.25}) will provide the link between the Schwinger
action principle and the Feynman path integral.

\section{\label{sec3}The Feynman path integral}

In order to solve Eq.~(\ref{2.25}), let
\begin{equation}
\ampJ=\int\left( \prod_i d\sigma^i( q_\ast;q)\right) F\left\lbrack
q_\ast;\sigma^i(q_\ast;q)\right\rbrack
\exp\left(-\frac{i}{\hbar}J_i\sigma^i\right) \;,\label{3.1}
\end{equation}
for some function $F$. The integration is assumed to extend over
all $\sigma^i( q_\ast;q)$ (or equivalently over all $q^i$, as will
be seen below) for which $\sigma^i( q_\ast;q) = \sigma^i(
q_\ast;q_1)$ at time $t = t_1$, and $\sigma^i( q_\ast;q) =
\sigma^i( q_\ast;q_2)$ at time $t = t_2$. If Eq.~(\ref{3.1}) is to
solve Eq.~(\ref{2.25}), we must have
\begin{eqnarray*}
0&=&\int\left(\prod_i d\sigma^i\right) \left\lbrace E_i\lbrack
q_\ast;\sigma^i(q_\ast;q) \rbrack -J_i\right\rbrace F
\left\lbrack q_\ast;\sigma^i(q_\ast;q)\right\rbrack
\exp\left(-\frac{i}{\hbar}J_i\sigma^i\right)\nonumber\\
&=&\int\left(\prod_i d\sigma^i\right) \left\lbrace E_i\lbrack
q_\ast;\sigma^i\rbrack F\left\lbrack q_\ast; \sigma^i\right\rbrack
+\frac{\hbar}{i}F\left\lbrack q_\ast;
\sigma^i\right\rbrack\frac{\delta }{\delta\sigma^i} \right\rbrace
\exp\left(-\frac{i}{\hbar}J_i\sigma^i\right) \;.
\end{eqnarray*}
If we integrate the second term in the last line by parts, then
\begin{eqnarray}
0&=&\int\left(\prod_i d\sigma^i\right) \left\lbrace E_i\lbrack
q_\ast;\sigma^i\rbrack F\left\lbrack q_\ast; \sigma^i\right\rbrack
-\frac{\hbar}{i}\frac{\delta F\left\lbrack q_\ast;
\sigma^i\right\rbrack }{\delta\sigma^i} \right\rbrace
\exp\left(-\frac{i}{\hbar}J_i\sigma^i\right)\nonumber\\
&&\left.+\frac{\hbar}{i}F\left\lbrack q_\ast;\sigma^i(q_\ast;q)
\right\rbrack \exp\left(-\frac{i}{\hbar}J_i\sigma^i\right)
\right|_{q_1}^{q_2}\;.\label{3.2}
\end{eqnarray}
Because $E_i=\delta\tilde{S}/\delta\sigma^i$, if we assume that
the surface term in Eq.~(\ref{3.2}) vanishes, then the solution to
Eq.~(\ref{3.2}) is
\begin{equation}
F\left\lbrack q_\ast;\sigma^i(q_\ast;q) \right\rbrack = f(q_\ast)
\exp\left(\frac{i}{\hbar}\tilde{S}(q_\ast;\sigma^i) \right)
\;,\label{3.3}
\end{equation}
for arbitrary $f(q_\ast)$. The condition for the surface term in
Eq.~(\ref{3.2}) to vanish is then that $S\lbrack q = q_1\rbrack =
S\lbrack q = q_2\rbrack$ (assuming that the source $J_i$ is only
non-zero for $t_1<t<t_2$.)  This condition is often met in field
theory by assuming that $q^i$ is in the vacuum state at $t = t_1$
and at $t=t_2$.

We have therefore found that
\begin{equation}
\ampJ=f(q_\ast)\int\left(\prod_i d\sigma^i\right)\exp\left\lbrace
\frac{i}{\hbar}(\tilde{S}-J_i\sigma^i)\right\rbrace \;.\label{3.4}
\end{equation}
The integration in Eq.~(\ref{3.4}) may be changed to one over the
more conventional variable $q^i$ as follows. The usual rule for a
change of variable gives
\begin{equation}
\left(\prod_i d\sigma^i(q_\ast;q)\right) = \left|{\rm
det}\,\frac{\delta}{\delta q^j} \sigma^i(q_\ast;q)\right|
\left(\prod_i dq^i\right)\;.\label{3.5}
\end{equation}
Noting from Eq.~(\ref{2.6}) that
$\sigma^i(q_\ast;q)=g^{ik}(q_\ast)\delta\sigma(q_\ast;q)/\delta
q_\ast^k$, and that the Van Vleck-Morette
determinant~\cite{VanVleck,Morette} is defined by
\begin{equation}
\Delta(q_\ast;q)=|g(q_\ast)|^{-1/2} |g(q)|^{-1/2} {\rm det} \left(
-\frac{\delta^2 \sigma(q_\ast;q)}{\delta q^i\delta
q_\ast^j}\right)\;,\label{3.6}
\end{equation}
Eq.~(\ref{3.5}) becomes
\begin{equation}
\left(\prod_i d\sigma^i(q_\ast;q)\right) = \left(\prod_i
dq^i\right) |g(q)|^{1/2}|\Delta(q_\ast;q)| |g(q_\ast)|^{-1/2}
\;.\label{3.7}
\end{equation}
Here $g(q)$ denotes ${\rm det}\,g_{ij}(q)$, and the factors of
$g(q),g(q_\ast)$ have been chosen to make $\Delta(q_\ast; q)$ a
scalar in each argument.

With the change of variable described above, the expression for
the transition amplitude becomes
\begin{equation}
\ampJ= |g(q_\ast)|^{-1/2}f(q_\ast)\int \left(\prod_i dq^i\right)
|g(q)|^{1/2}|\Delta(q_\ast;q)| \exp\left\lbrace
\frac{i}{\hbar}(\tilde{S}-J_i\sigma^i)\right\rbrace \;.\label{3.8}
\end{equation}
The amplitude must be invariant under the change of coordinates
$q_\ast^i\rightarrow q_\ast^{\prime i}$. This is seen to constrain
$|g(q_\ast)|^{-1/2}f(q_\ast)$ to transform like a scalar. This
scalar is irrelevant since we typically only compare one amplitude
with another. In fact, if we require the expression to reduce to
that of Feynman when the space is flat, then
$|g(q_\ast)|^{-1/2}f(q_\ast)$ must be a constant. In any case,
because $|g(q_\ast)|^{-1/2}f(q_\ast)$ has no dependence on the
dynamical variables $q^i$, we may simply take
\begin{equation}
\ampJ= \int \left(\prod_i dq^i\right)
|g(q)|^{1/2}|\Delta(q_\ast;q)| \exp\left\lbrace
\frac{i}{\hbar}(\tilde{S}-J_i\sigma^i)\right\rbrace \;,\label{3.9}
\end{equation}
as the path integral representation for the amplitude. As promised
in the introduction, in addition to the natural volume element
$\displaystyle{\left(\prod_i dq^i\right) |g(q)|^{1/2}}$, the
additional factor of $\Delta(q_\ast;q)$ is seen to be present. As
we have already mentioned, this agrees with the conclusion of
Ref.~\cite{stochastic} and as shown by Parker~\cite{Parker}
results in a Schr\"{o}dinger equation with no explicit dependence
on the scalar curvature.

\section{\label{sec4}Discussion and conclusions}

In addition to the natural volume element in the path integral
measure, we have shown that there is an additional term which
involves the Van Vleck-Morette determinant. The origin of this
term can be traced to consistency between the Schwinger action
principle and the Feynman path integral. In the special case of a
flat space, $\Delta(q_\ast;q)=1$, so that this additional term
disappears even if curvilinear coordinates are used. As mentioned
in the introduction, the existence of the term
$|\Delta(q_\ast;q)|$ in the measure leads to the normal
Schr\"{o}dinger equation without any additional modifications due
to the curvature~\cite{Parker}.

There are of course other ways to derive the factor of
$\Delta(q_\ast;q)$ in the measure. One is the previously mentioned
method of DeWitt-Morette et al.~\cite{stochastic}. Another
approach, which is independent of the Schwinger action principle,
is to postulate that the amplitude satisfy the equation of motion
Eq.~(\ref{2.25}). This is what would be done following
Symanzik~\cite{Symanzik} for example. The steps leading up to the
end result of Eq.~(\ref{3.9}) are identical. We chose instead to
postulate the more general action principle of Schwinger, and to
derive Eq.~(\ref{2.25}) as one of its many consequences.

It is of interest to explore the consequences of the measure found
in this paper in the case of quantum field theory. A covariant
approach to quantum field theory has been advocated by
Vilkovisky~\cite{Vilkovisky}. It would be of interest to study the
implications for gauge theories, particularly in relation to the
geometrical analysis of the measure presented in
Refs.~\cite{EKT1,EKT2}.

\end{document}